# Reference-free dual-comb spectroscopy with inbuilt coherence

MIKHAIL ROIZ[1,\*,+], SANTERI LARNIMAA[1,\*], TOUKO UOTILA[1], MIKKO NÄRHI[2] AND MARKKU VAINIO[1,2,3]

[1]*Department of Chemistry, University of Helsinki, FI-00560, Helsinki, Finland*
[2]*Photonics Laboratory, Physics Unit, Tampere University, Tampere, FI-33101, Finland*
[3]*e-mail: markku.vainio@helsinki.fi*
[\*]*These authors contributed equally to this work*
[+]*Corresponding author: mikhail.roiz@helsinki.fi*

We demonstrate a simple system for dual-comb spectroscopy based on two inherently coherent optical frequency combs generated via seeded parametric down-conversion. The inbuilt coherence is established by making the two combs share a common comb line. We show that the inbuilt coherence makes it possible to use a simple numerical post-processing procedure to compensate for small drifts of the dual-comb interferogram arrival-time and phase. This enables long-time coherent averaging of the interferograms.

Recent progress in Optical Frequency Comb (OFC) technology has unleashed the full potential of high-precision spectroscopy [1]. One of the techniques that made a major contribution to it is dual-comb spectroscopy (DCS) [2, 3]. DCS is an interferometric method that enables fast broadband measurements without any mechanical scanning parts [4]. The interferograms form as a consequence of asynchronous sampling of one comb by the other comb with a slightly different repetition rate. Once the interferograms are recorded with a photodetector and a digitizer, they can be Fourier transformed to obtain the spectrum.

In DCS, long-time coherent averaging relies on the mutual coherence between the two OFCs. This is usually achieved through intermediate reference lasers along with phase-locking loops, which is a complication for systems that target field applications. An alternative way to enable coherent averaging is the use of adaptive sampling [5]. Here, the signals that represent the mutual fluctuations between the two free-running OFCs are extracted through the introduction of intermediate reference lasers. These signals are then used to correct the interferogram phase and resample the interferograms thus eliminating incoherence between the combs. However, this approach is quite tedious to implement in the most general case of two free-running OFCs. Hence, the inbuilt or passive coherence is of a great interest because it simplifies the DCS setup in many ways.

Recently, there have been demonstrations of various DCS platforms with inbuilt coherence [6-9]. One such approach is based on a single mode-locked laser that emits two OFCs with different repetition rates from the same cavity, hence the name – dual-comb lasers. Since both combs share the same cavity, they are highly coherent with one another leaving only small relative fluctuations of the offset frequency and repetition rate [9].

In this paper we present an alternative and simple method for generating mutually coherent OFCs in the near- and mid-infrared regions simultaneously. The method is based on a single-pass continuous-wave (CW) seeded femtosecond optical parametric generation (OPG) [10-13]. We have recently shown how this method can be used to generate fully stabilized and highly versatile mid-infrared frequency combs in a simple configuration [14]. The method is very efficient thanks to the pulse trapping effect, which forces the pump, signal and idler pulses to propagate with the same velocity in a nonlinear crystal [15, 16]. The key feature of the method is that the seed laser acts as a first pre-generated comb line in the signal or idler region. This is a great advantage for DCS because one can use the same seed laser for two CW-seeded OPG-based OFCs. In this configuration, the shared seed eliminates the carrier-envelope offset (CEO) difference between the two combs ($\Delta f_{CEO} = 0$) almost completely. Fluctuations in the second degree of freedom – namely the repetition rate difference ($\Delta f_{RR}$) – can be easily compensated by a simple post-processing algorithm. Here we demonstrate that such an algorithm exhibits an excellent performance allowing for comb line resolved DCS without using reference lasers. This is done by comparing the measurement results for two different configurations: (1) when the adaptive sampling technique with CW reference laser is used to compensate for the repetition rate difference drifts, (2) when the numerical correction algorithm that does not require any reference lasers is used instead.

In Fig. 1a one can see a schematic that describes the basic principle of the spectrometer. The experimental details of CW-seeded OPG can be found in our previous articles [10, 11].

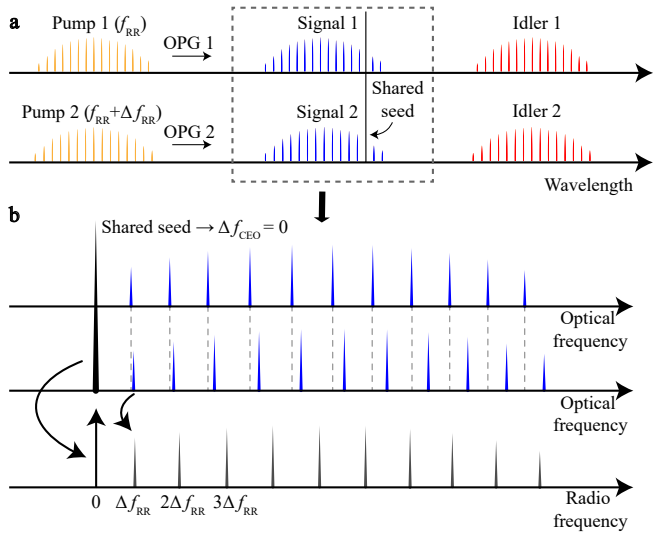

Fig. 1. (a) The dual CW-seeded OPG setup for DCS. (b). Schematic of the DCS radio frequency spectrum formation.

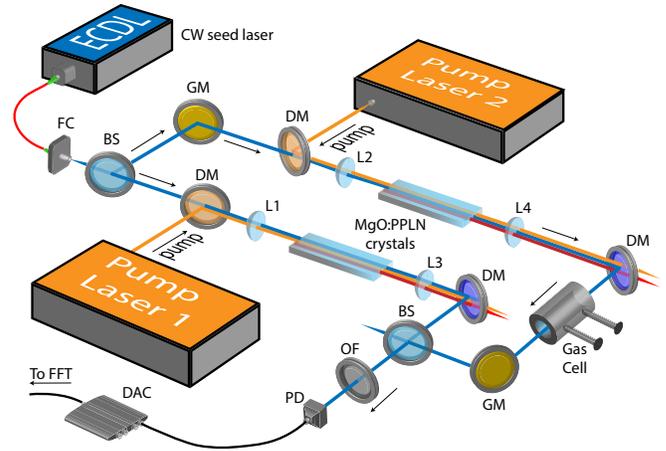

Fig. 2. Simplified schematic of the DCS setup. OF: optical filter, GM: gold mirror, DAQ: data acquisition card, BS: beam splitter, FC: fiber collimator, DM: dichroic mirror, L: lens, PD: photodiode, ECDL: External Cavity Diode Laser, FFT: fast Fourier transformation.

Two free-running femtosecond lasers (MenloSystems GmbH, 1042 nm central wavelength emitting 100 fs pulses at 250 MHz repetition rate, <200 kHz linewidth in 100 ms) drive the pulse-trapped OPG processes in two periodically-poled lithium niobate (PPLN) nonlinear crystals, where signal and idler frequency combs are produced with 1538 nm and 3380 nm central wavelengths, respectively. The output beam of a CW laser (Toptica Photonics, CTL 1550, <1 kHz instantaneous linewidth) operating at 1532 nm is evenly split and sent to each crystal. Depending on the wavelength of seeding, one can establish coherence between the two signal or idler combs [12, 13]. In this work, we choose to work with the signal combs, since it is easier to characterize the system in that spectral range. The pump lasers, and thus the respective signal and idler combs, have slightly different repetition rates of $f_{RR}$ and $f_{RR}+\Delta f_{RR}$. The CW seed laser serves as a common comb line between the two signal combs mapping the first beat note radio frequency (RF) to zero Hertz. Consequently, each pair of comb lines maps the optical frequencies to RF corresponding to $\Delta f_{RR}$, $2\Delta f_{RR}$, $3\Delta f_{RR}$ and so on, as schematically exemplified in Fig. 1b.

To measure absorption spectra of gases, first one needs to measure the interferograms. This is done by combining the generated signal combs using a 50:50 plate beam splitter and sending them to a photodiode as shown in Fig. 2. One of the combs passes through a gas cell, while the other comb is simply sent to the detector. The two combs interfere due to the difference in repetition rates, producing interferograms. The interferograms are then digitized using a fast data acquisition card (DAQ, GaGe CompuScope 142000, 14 bits, 200 MS/s) and Fourier transformed to obtain the desired RF power spectrum.

First, we studied how well the passive coherence is established via the shared CW seed laser. Since we use independent free-running pump lasers, relative timing jitter between the two combs leads to changes in $\Delta f_{RR}$. Hence, for the initial characterization of our system, we optically locked $\Delta f_{RR}$ by using a phase locked loop. This is done with the help of a CW reference laser introduced in the OPG signal region but shifted away from the seed laser wavelength. We separately produce beat notes of this laser against each signal comb and combine them using an RF mixer. With the assumption that the CEO difference between the combs is eliminated via the shared seed laser, the resulting mixer signal indicates only the mutual repetition rate fluctuations between the two combs. This signal can be used for phase locking. We will refer to this signal as the reference signal. More details about the reference signal can be found in Supplement 1.

We optically locked $\Delta f_{RR}$ to 11.7 kHz using the reference signal and measured a set of interferograms during an ~80 ms time window. This resulted in a total number of 937 interferograms. The time window of 80 ms will be referred to as a segment further in the text. The relative comb linewidth is 12.5 Hz, which is limited by the measurement time of the segment. When $\Delta f_{RR}$ is optically locked, we can measure interferogram phase fluctuations independently of $\Delta f_{RR}$. In Fig. 3a one can see an example of real measured interferograms at different times during a segment. The first interferogram triggers the DAQ card and it is clear that the shapes of the subsequent interferograms slightly differ from the triggering one. To represent the drift in numbers, we deduced the interferogram phase using a phase retrieval algorithm. The information on the phase as a function of time could be used to introduce a time dependent phase shift within the segment to make the interferogram shapes appear the same before Fourier transformation (see Supplement 1 for details). From Fig. 3b it is clear that the phase change is small (well below $\pi/8$), for which reason we did not perform the phase correction within one segment. In our setup, the interferogram phase fluctuations appear due to the interferometric path length fluctuations of the seed when it is split into two arms. This was confirmed by placing a mirror with a piezo element in one of the arms and sending a periodic signal to the piezo. As a result, the interferogram phase changed periodically.

The reference signal can be used directly for the adaptive sampling technique, which helps to avoid phase-locked loops. Although the technical implementation of the optical locking and adaptive sampling differs, their outcome is essentially the same. In our next experiments we emulated the adaptive sampling technique by digitizing the reference signal (along with the segment data) and using it to resample the interferograms afterwards. The same segment data was processed with our numerical algorithm without using the reference signal. Hereafter, the results of our numerical algorithm will be compared to those obtained from the emulated adaptive sampling technique. For the comparison, we calculated how much the arrival time of interferograms deviates from the linear trend using the two methods mentioned above (see Fig. 3c). In other words, we deduced the degree of nonlinearity in the time axis during one segment due to $\Delta f_{RR}$ changing in time. It is clear that the numerical algorithm is capable of retrieving the arrival-time deviation correctly without relying on any CW reference lasers. Once the arrival-time deviation is deduced correctly, this information can be used to linearize the time axis and resample the segment data. After resampling, a full segment of 937 interferograms is Fourier-transformed. If several segments are measured, the Fourier transform is carried out after coadding the phase-corrected segments (see Supplementary 1 for details).

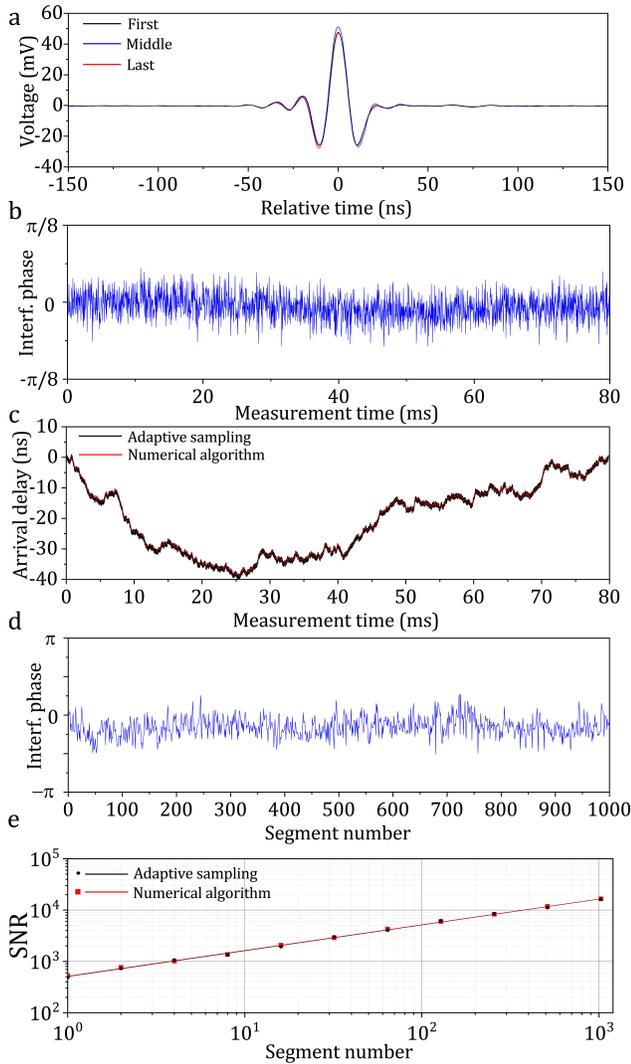

Fig. 3. (a) The triggering and subsequent interferograms in one segment (80 ms time window containing 937 interferograms). (b) The interferogram phase fluctuations within one segment retrieved using the phase retrieval algorithm. (c) Arrival-time deviation of interferograms in one segment. (d) The interferogram phase fluctuations in between segments. (e) Spectral SNR as a function of segment number.

Next, we performed the long-time averaging experiment to see whether our numerical method gives a reasonable result. We recorded 1024 segments and deduced the arrival-time deviation in each of them using the same two methods. In addition, the phase unwrapping algorithm was used to determine the interferogram phase drift in between the segments (see Fig. 3d). We assumed that the interferogram phase is constant within one segment and calculated it only for the first interferogram in each segment. The determined interferogram phase (see Fig. 3d) along with the arrival-time deviations were used to correct the interferograms before coaddition. The same mutual phase correction between segments was also implemented for the adaptive sampling case. The dead time between each segment was ~1 s, which is the reason why the interferogram phase drifted more than shown in Fig. 3b. The total measurement time including the dead time was about 19 minutes, and the effective measurement time was 82 s, resulting in the coherent averaging of almost 1 million interferograms. To compare the two methods, we calculated how the signal-to-noise ratio (SNR) of spectra behaves as a function of measurement time. The SNR is calculated by taking the maximum value of the RF comb tooth peak and dividing it by standard deviation of the noise around the peak. The result can be seen in Fig. 3e. The SNR increases linearly with the number of averaging segments. Both plots have a slope close to 0.5, indicating that the noise averages out as white noise. This confirms that the long-time coherent averaging performance is equivalent in the two cases under consideration.

Last but not least, we used the above-mentioned long-time averaging scheme for spectroscopic measurements. We filled the gas cell with 6 % acetylene at 300 mbar total pressure. In Fig. 4 one can see an example of the acetylene spectrum averaged for the effective time of 82 s and the result of fitting according to HITRAN database [17, 18]. We also calculated the residuals between the spectrum obtained using our numerical algorithm and the HITRAN fit as well as between the adaptive sampling approach and its HITRAN fit (see Fig. 4). The standard deviation of the residuals is about $9.9 \times 10^{-4}$ and the standard deviation of the difference between residuals is $9.2 \times 10^{-5}$. This proves that the two methods essentially lead to the same spectroscopic results. Using the method described in [19], we calculated the figure of merit to be $6.9 \times 10^6$.

In conclusion, we have demonstrated a simple DCS setup with inbuilt coherence based on two CW-seeded OPGs. The mutual coherence is established via a shared CW seed laser, which greatly simplifies the experimental setup. Since the two frequency combs are inherently coherent, only minor adjustments due to drifts in the repetition rate and offset frequency differences need to be performed using our numerical correction algorithm. We proved that the numerical algorithm correctly determines the arrival-time deviations caused by drift of $\Delta f_{RR}$, thus enabling coherent averaging without relying on any special reference signals. We anticipate that the presented DCS method along with the developed numerical algorithm could open up opportunities towards efficient and completely free-running DCS systems also in the mid-infrared spectral region. Our system is well suited for applications that do not require real-time spectroscopic acquisition, otherwise the adaptive sampling can be used instead [12, 13, 20]. In addition, we believe that the rapidly developing dual-comb laser technology [9] can complement our method. If used as the pump source, this can further minimize the requirements for repetition rate corrections making the coherent averaging very time efficient.

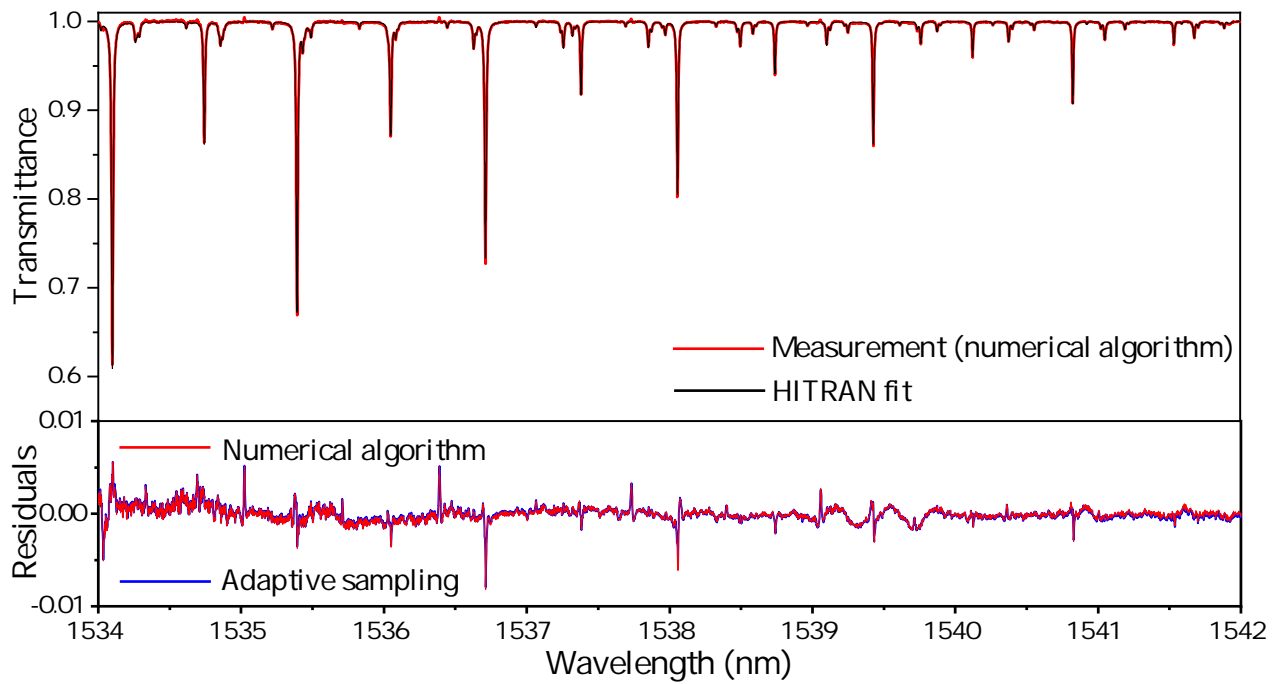

Fig. 4. Acetylene spectrum obtained by coherently averaging 1024 segments (each containing 937 interferograms). The bottom panel shows the residuals between the spectrum obtained using the numerical method and its corresponding HITRAN fit (red) and the residuals between spectrum obtained with the adaptive sampling and its corresponding HITRAN fit (blue). The spectral point spacing of the measurement is 250 MHz.

**Funding.** Alfred Kordelin Foundation; University of Helsinki; European Partnership on Metrology (22IEM03 PriSpecTemp); European Union's Horizon Europe Research and Innovation Programme.

**Acknowledgment.** We thank Dr. Juho Karhu from the University of Helsinki for fruitful discussions. Santeri Larnimaa acknowledges financial support from the CHEMS doctoral program of the University of Helsink. The project (22IEM03 PriSpecTemp) has received funding from the European Partnership on Metrology, co-financed from the European Union's Horizon Europe Research and Innovation Programme, and by the Participating States

We thank Dr. Juho Karhu from the University of Helsinki for fruitful discussions.

**Disclosures.** The authors declare no conflicts of interest.

**Data availability.** Data presented in this paper are available from the authors upon reasonable request.

**Supplemental document.** See Supplement 1 for supporting content.